\documentclass[12pt]{article}
\usepackage{graphicx}
  \usepackage{color}
\usepackage{epsfig}
\usepackage{amsmath}
\usepackage{hyperref}
\begin{document}

\begin{centering}
\title{ 
{\color{black}  Hierarchy Selection: New team ranking   indicators for cyclist  multi-stage races}
}
\vskip0.5cm
\author{ Marcel Ausloos$^{a,b,c,}$}

\vskip0.5cm

$^a$ School of Business, University of Leicester,\\
Brookfield, 
Leicester, LE2 1RQ, UK\\   e-mail: ma683@leicester.ac.uk\\
 $^b$ Department of Statistics and Econometrics,  \\ Bucharest University of Economic Studies,  15-17 Dorobanti Avenue, \\ District 1, 010552, Bucharest, Romania, \\  e-mail: marcel.ausloos@ase.ro
 \\ $^c$ Group of Researchers Applying Physics in Economy and Sociology \\(GRAPES),  Beauvallon, rue de la Belle Jardini\`ere, 483/0021\\ Sart Tilman, B-4031, Li\`ege Angleur, Belgium, Europe  \\
  e-mail: marcel.ausloos@uliege.be

\end{centering}
\maketitle
\newpage 
 
\begin{abstract}

In this paper,  I report some investigation discussing  
{\color{black}  team selection, whence hierarchy, through} ranking indicators, for  {\color{black} example when}  “measuring" professional cyclist team's “sportive value", in particular in multistage races. 
A logical, it seems,  constraint is introduced on the riders: they must finish the race. Several new  indicators are defined, justified, and compared.  These  indicators are mainly based on the arriving place of  (“the best 3") riders instead of their  time needed for finishing the stage or the race, - as presently classically used. A case study, serving as an illustration containing the necessary ingredients for a wider discussion, is the 2023 Vuelta de San Juan, {\color{black} but without loss of generality}. 
{\color{black} 
It is shown  that the new  indicators offer some new viewpoint for distinguishing the ranking through the cumulative sums of the places of riders rather than their finishing times.  On the other hand, the  indicators indicate a different team hierarchy if only the finishing riders are considered. Some consideration on the “distance" between ranking  indicators is presented.}
Moreover, it is argued that these new ranking indicators should hopefully promote more competitive races, not only till the end of the race, but also until the end of each stage. Generalizations and other applications 
{\color{black} within operational research topics, like in academia,}  are suggested.
 
\end{abstract}

   
\maketitle
 
\vskip0.4cm
\newpage
 
 \section{Introduction}\label{Introduction}
 
  {\color{black}  Operational researchers mostly focus on how to help organisations develop better business systems, to examine how an organisation operates and to suggest more effective ways of working, through individuals and procedures. Let it be called  a “microscopic approach” (Ackerman et al., 2018). In contrast, operational research (OR)  on teams (scoring and ranking) looks like a “mesoscopic approach” (Corvalan, 2018). That should be part of the modern pillars of OR, within collective choice frameworks research and applications. Indeed, 
 ranking   is common in many social life activities: politics, media, economics, academia, - and sports.
 
 It is of common knowledge that  for our planet evolution, optimizing (or optimized) selection of choices are mandatory (Lamarck, 1815-1822; Darwin, 1859; 
  Heider, 1958; Ebeling  and Feistel, 2011). Both endogenous and exogenous criteria must be provided for optimizing  choices among possibilities (Dyer and Miles, 1976; Csató, 2021). Much research has been done on pair competition, as in Verhulst-prone scenarios, through prey-predator or epidemic models (Vitanov et al., 2010; Caram et al., 2015;  Saeedian et al., 2017; Gawroński et al., 2022), or for multi-agent interactions (Lambiotte and Ausloos, 2007; Csató, 2020, 2021, 2022).  Sometimes, the final choice, when reduced to comparing pairs, lead to paradoxical situations (Condorcet, 1785; Arrow, 1950; Bozóki et al., 2016; Ágoston and  Csató, 2022).   
  Moreover, the order of criteria might lead to ambiguities (He and Deng, 2023). In brief, one may recall trivial methods of
  preference aggregation techniques, like  the   “means of scores",  - weighted or not, the  “Maximum Likelihood Rule" (Le Cam, 1990; Varela and Rotundo, 2016),  based on the concept of   pairwise preference notions, and TOPSIS, - a method of compensatory aggregation that compares a set of choices, identifying weights for each criterion, normalising scores for each criterion, and calculating the geometric distance between each choice and the ideal one, which is the best score in each criterion  (Yoon, 1997; Hwang et al., 1993; Lai et al., 1994). 

The discussion, and the subsequent conflict resolution,  pertains to a comparison of the evaluation methods, according to criteria  (Krawczyk et al., 2019; Krawczyk  and Ku{\l}akowski,  2021). 
The drastically annoying deduction seems to stem from the plethora of  “preference parameters", i.e., thus possible criteria. Practically, one turns toward aggregation processes (Munda, 2012), from multi-dimensions toward a single number. The complexity matter further arises when the hierarchy selection is depending both on individual  (leadership, PI,...) and on a team (past and present members) achievement scoring. 
 }

  There are many papers published on   “ranking teams",  {\color{black} e.g., among pioneers, Sinuany-Stern (1988), Churilov and Flitman (2006), and  Dadelo et al., (2014),}  although many less than on  {\color{black} leaders or} athletes rankings. {\color{black} However, an objective “team value" hierarchical scoring is of high relevance for multi-teams collaborations and competitions. Much complexity is known to exist (Fishburn, 1981; Churilov and Flitman, 2006), like in soccer (Ausloos, 2014; Ausloos et al., 2014a, 2014b; Csató, 2020; Ficcadenti et al., 2023); } 
  {\color{black} for a recent literature review of contemporary interest on cycling, see Van Bulck et al. (2023). }
   Such considerations are emphasized  on one  sport event, in the present study, for justifying arguments and subsequent empirical analysis.

 The largest amount {\color{black} of papers on team ranking} pertains to the most popular sports, of course, like (American) football, soccer, and basketball. Many methods have been proposed.  A modern  overview by Sorensen (2000) 
   points to various methods for ranking, but mainly pertinent for team duels competitions.
  Sorensen (2000) 
 and  Vaziri et al.  (2018) 
  appear to be relevant, but discuss concepts rather than applications.     {\color{black} More recent considerations can be found in  Csató (2017a, 2017b, 2020, 2021, 2023) and Kondratev et al. (2023).}  About  cycling teams races and subsequent  ranking, the literature is much less abundant{\color{black}; yet, recall  Van Bulck et al. (2023).}
  
  Based on such considerations, some research could be undertaken in finding a new, non classical, way of ranking (professional) cyclist teams. 
 {\color{black} {\it Mutatis mutandis}, several aspects can be diverted toward (team) ranking in other social life activities, e.g., in academia or marketing,  where some preference scoring, whence some hierarchy, is mandatory. Complex aspects have attracted attention, as in Jose et al. (2008). 
 }
 
 Let us {\color{black} specifically} consider multi-stage races by professional cyclists. The most famous, prestigious,  ones are the   “Tour de France",   “Giro d'Italia",    “Vuelta a España".  Beside these   “Grand Tours", there are many others (so called Elite, 2.HC Stage Races,  within UCI ProSeries) to which the present considerations apply.
 
 Even though such cyclist races are won by one rider,  the role of the team is of crucial importance (Albert, 1991; 
{\color{black} Mignot, 2015, 2016}; Cabaud, 2022). After each stage, a team ranking is provided by the race organisers, according to UCI rules\footnote{\url{https://www.uci.org/regulations/3MyLDDrwJCJJ0BGGOFzOat}}.  The teams are ranked according to the aggregated finishing time of the fastest 3 riders of a team for that stage, - excluding all so called bonus time.
  That sum is cumulated after each stage. At the end of the multi-stage race,  each final team time is the result from the sum of such stage times, irrespectively of the involved riders.
 
 Let the finishing time of these 3 fastest riders, $i=1,2,3$, in their arriving  order, of team ${(\#)} $, for a stage $s$, be defined as  $ t^{(\#)}_{i,s}$.
 In mathematical terms, one calculates the   “team (finishing) time  on  stage $s$" as 
 
 \begin{equation}\label{tsteameq}
t_s^{(\#)}  = \Sigma_{i=1}^{3} \;\;  t^{(\#)}_{i,s}\;.
\end{equation}

 At the end of  a  $L$ stages race, one obtains each team $  {(\#)} $   “finishing time"  $   T_L^{(\#)}$ from the sum of each stage   “team time":

\begin{equation}\label{Tteameq}
  T_L^{(\#)}  = \Sigma_{s=1}^{L} \;\; t_s^{(\#)}  \;.
\end{equation}

Notice that it is often occurring that a race ends with a sprint by a huge bump of riders, thus all such riders are supposedly arriving at   “the same time" as the winner.
 When teams finish with an equal time, they are distinguished, whence  ranked, according to the sum of the 3 places of the relevant riders. Let such riders be at place  $ p^{(\#)}_{i,s}\;$,  with $i$=1, 2,  3.
Similarly  to the above, one can define the    “more objective" {\it team ranking place} $p_s^{(\#)}$ as resulting from 

 \begin{equation}\label{psteameq}
p_s^{(\#)}  = \Sigma_{i=1}^{3}  \;\;  p^{(\#)}_{i,s}\;,
\end{equation}
on stage $s$,  
and  calculate some 
\begin{equation}\label{PLteameq}
  P_L^{(\#)}  = \Sigma_{s=1}^{L}  \;\; p_s^{(\#)}  \;.
\end{equation}
at the end of the multi-stage race, for the final ranking, - according to the team placing at different stages, again irrespectively of the involved riders.

 Notice that both $t$ and $p$ lists do not necessarily give the riders in the same order, due to the last (3) kilometre(s) neutralisation rule, allowing riders to have   “technical problems", tire punctures, even falls, or willingly stop racing,  
 along such a distance.

Arithmetically,  within these measures,  Eqs.(1)-(4), it can still occur that some teams may have an equal rank at the end of a given stage, - as well as later on at the end of the race, -  if the respective sums allow so.  (In this study, it is considered that an {\it ex aequo} should remained to be what it is, the same rank, without adding a new criterion.)
 
  Two questions seem to arise
  \begin{itemize}
  \item $Q_1$: 
  why should one conclude at the end of a multi-stage race that the   “final team ranking" results from the time (or place) of riders who do  {\it not even} finish the race?
  \item $Q_2$: 
  why should one prefer  the    $T_L^{(\#)} $ to the   $P_L^{(\#)}$  measure?
  \end{itemize}
  
 On $Q_1$:   it is bizarre that the   “final congratulations" about being the   “winning team" is based on missing riders at the end of a race. One may admit that some team strategy, based on a rider specific prowess for a given stage,  might make sense for specific stages on specific days, but it seems that an objective measure should only consider those riders who finished the race.
  
 For example, admitting this consideration, one   avoids a case like that of M. Cipollini in the Tour de France 1999: he won 9 stages, among them were 4 consecutive  (“flat'') stages,  but he    left  (abandoned) the race on the next rest day, when stages were reaching mountain climbing features. Yet, his winning times contributed to the final   “team total time" measure.
 
   In a sophist way, one could then suggest that a team could  bring up new riders in a multi-stage race, replacing athletes at will, as done in basketball, hockey,  ... or as in other sports involving team duels.
 
 On $Q_2$, recall that  as far as, e.g.,  the 1905 Tour de France, the race winner rider was not deduced from the aggregated  time-based  system, but from a place-based system. This system lasted until  the  1913 race, when the time-based system was re-introduced. Even though the change from the original (1903)  time-based counting was due to some scandal (the time-based winning rider was   accused to have been transported by car or rail during a stage),  this change in  scoring demonstrates that the  {\it rider} place-based system can be less  easily   “adapted" to potential cheating than a time-based system. The more so, thereby I argue, for  {\it team} based ranking:   manipulation,  place rigging through trading and/or biasing the final count,  seems  more easily avoided\footnote{Several riders prepare the sprint for their best sprinter, a so called train, each   lead out   “locomotive  rider" successively peeling off in turn in order to hopefully give their fastest sprinter the best opportunity to win. The accounted time of the train riders becomes weakly relevant, as long as they remain in the bunch during the last 3 kilometres. I argue that the place accounting should demand more riding action till the end of a stage, thus more objective  competition. }.
 
 Moreover, one avoids the paradoxical situation, mentioned here above, that, in order to rank teams without {\it ex aequo}, one imposes the   “place filter" onto the   “time filter".  Of course, {\it ex aequo's}  can still exist, after summing 3   “small" integers,  in Eqs.(\ref{psteameq})-(\ref{PLteameq}).
 
  Thus, both for answering  $Q1$ and   $Q_2$,
  it seems that one can argue that riders should not give up too early in the race,  but should keep up in order to be among their team 3   “best" riders, whatever their skill. Indeed, a   “good place" on the arrival line is far from having to be disregarded, with these new measures.  
  
In the same line of thought,  it is argued that an appropriate measure team's value, and their ranking,  should correspond to the remaining  riders of each  team at the end of the multi-stage race. 
   
 For further numerical display and discussion, it is appropriate to select some case studies;  the data used in this study is taken from a recent race,  but without any lack of generality.
  
 {\color{black} 
It is shown and concluded  that the new  indicators offer some new viewpoint for distinguishing the team rankings through the cumulative sums of the places of riders rather than their finishing times.  On the other hand, the  indicators indicate a different team hierarchy if  the finishing riders are those specifically taken into account. Some consideration on the “distance" between  indicators is presented.
}

 The rest of the paper content goes as follows.  First, the data used in this study, taken from a recent race, is briefly described in Sec. \ref{Data},

  In Sec. \ref{finalteamtime},  two  indicators are proposed based on the   {\it stage finishing time} of cyclists  {\it finishing} a multi-stage race.

 In Sec. \ref{teamfinalplace},  two  indicators are proposed based on the   {\it finishing place} of cyclists    {\it finishing} a multi-stage race.
 
  In Sec. \ref{overallbestriderplaceemphasis}, two other new indicators are proposed based on the finishing place of cyclists in multi-stage races, bearing a different emphasis on the riders finishing places in the various stages.
 
    In Sections  pertaining to some   “analysis",    the statistical discussion of results is based on the Kendall-$\tau$ coefficients classically used for comparing ranks in equal size  lists ({Kendall, 1938; Abdi, 2007; Puka, 2011).  {\color{black} Due to a reviewer comment, a brief set of considerations on improving the Kendall $\tau$ usage is found in Appendix A, based on  “weighted preferences'' notions as discussed by Can (2014).}

In Sec. \ref{Discussion},  a discussion of the methodology and data  analysis is  followed by conclusions in Sec. \ref{Conclusions}; the latter  contains {\it a posteriori} arguments in favour of the new  indicators,  with suggestions for further research and applications in organizations confronted to a selection process based on a hierarchy list.
 
 
 Moreover, in addition, for later discussion here below, it  is useful to introduce some  notations in order to distinguish individual riders: each is attributed  a   “bib number" $d$ by the race organizer. Thus, the usually recorded time necessary for the  rider  $d$ to finish a given stage $s$ is thereafter noted $t_s^{(d)}$, while its finishing place on the  stage $s$ is  called $p_s^{(d)}$. 
  When  the team and specific rider have both to be emphasized, one will note  such  measures as    $t_{d,s}^{(\#)}$   and $p^{(\#)}_{d,s}$  respectively.
 
  Recall that at the end of the $L$-stage race,  riders are hierarchically ranked, according to   UCI rules, along their  aggregated time, here called $T_L^{(d)} =  \Sigma_{l=1}^L \; \; t_s^{(d)} $, in ascending order, which leads to their final (time) ranking. The latter should be distinguished from    $P_L^{(d)}=  \Sigma_{l=1}^L \; \; p_s^{(d)} $, resulting from merely summing all arriving places of a given $d$ rider in the various $s$ stages.
 
 \section{Data} \label{Data}  
 
 For the present purpose, to deal with data on long (3 week) races would burden the discussion, and would not give much  more weight to findings, whence arguments. 
 
  Therefore, the case study is the recent   2023  Vuelta Ciclista a la Provincia de San Juan: \url{https://www.vueltaasanjuan.org}.  The  
  race is part of the UCI ProSeries calendar in category $2.Pro$.
  The 2023 race, in brief thereafter called VSJ,  took place over ($L=$) 7 stages from (Sunday) Jan. 22 till  (Sunday) Jan. 29; there was a one day rest on (Thursday) Jan 26.   There were ($M=$) 26 teams: 25 with ($n=$) 6  starting riders, and  one only with 5 starting riders. 
 Thus, the initial bunch was ($nM-1=$) 155 rider wide; only 133 riders finished the race, due to abandoning or ''arriving after delays'' cyclists.  There were 7 WorldTeams, 5 ProTeams, 10 Continental teams and 4 National teams. For each team, the official UCI code is hereby used, - for shortening the writing and avoiding undue publicity claims.
 
 
 The interesting data is obtained from  the chronometer officials websites
\url{https://www.edosof.com/carrera/517/clasificacion/11\#cajaetapas}, and similar web sites for the different stages. One can  notice that a direct access to the 3rd stage results is missing, i.e.,
\url{https://www.edosof.com/carrera/517/etapas/31$\#$cajaetapas}, is empty, but these can be obtained from another media website, like
  \url{https://www.esciclismo.com/actualidad/carretera/74160.html}.
The 4th stage data is also misreported, i.e.
 \url{https://www.edosof.com/carrera/517/etapas/41$\#$cajaetapas}, 
 but the clumsiness can be  immediately resolved, and cross checked  through media websites, like
  \url{https://www.esciclismo.com/actualidad/carretera/74177.html}.
 
 
A warning: other websites also contain   “errors"; for example, not giving the full list of  finishing riders.  However, any missing information was manually supplemented, and cross checked through different websites. See also some official errors or misprints, described in the Appendix  {\color{black} B}, but not leading to consequences on the {\it teams place ranking relative values} which is the kern of this study.
 
 {\color{black}  To be more precise and for completeness, the used data is found in Supplemental Materials.}

\section{Team  Final  Time}    \label{finalteamtime} 

 The classically reported team final time (excluding time bonuses) is $T_L^{(\#)}$, as defined in Eq.(\ref{Tteameq}). The $T_L^{(\#)}$ of the 26 teams at the end of the VSJ multi-stage race, as so classically determined,  is given  in Table \ref{Table1TAVSJ}{\color{black} ; the conventional UCI acronyms are recalled in Table \ref{TableVSJacron}.}

 As explained here above, one can define the  thereafter called   “adjusted team final time", $ A_L^{(\#)} $, calculated as follows 
 
 \begin{equation}\label{Ateameq}
  A_L^{(\#)}  = \Sigma_{j=1}^{3} \; \; t_{j,L}^{(\#)}
\end{equation}  
where, in Eq.(\ref{Ateameq}),  $j$ = 1, 2, 3 refers to the   “3 best", whence fastest,  riders of the team ${(\#)}$  {\it having  completed all}  $L$ stages.  Thus,  $A_L^{(\#)}$  can only be so obtained at the end of the multistage race.  Yet, some possibility exists to   “generalize" the concept and its application; see the conclusion Section on    “further research" suggestions.
 Let it be emphasized that these 3   “$j$" riders  {\it might} be quite different from the various 3   “$i$" riders having contributed to any  $t_s^{(\#)}$, 
whence to   $T_L^{(\#)}$.   

The results pertinent for 2023 VSJ are given in Table \ref{Table1TAVSJ}, where one can be comparing  $T_L^{(\#)}$, not only with the officially published  $[T_L^{(\#)}]$ results, but also $A_L^{(\#)}$; recall to read the  comment on data misreporting in the Appendix  {\color{black} B}, leading to the   “definition" of             $[T_L^{(\#)}]$.
Notice at once that the   “best team" is markedly different in both measures: IGD is loosing its  first rank for MOV, - for about 40 sec. The largest move up concerns TFS; the biggest fall is for TBG. The respective order of other teams is also quite scrambled. 

From a statistical difference perspective, the Kendall $\tau$ Rank-Rank correlation coefficient	is equal to 0.79692 (Score = 259; Denominator = 325)\footnote{Conventionally, the number of concordant pairs is called $C$; that of discordant pairs is $D$. The    “score" $S$  is  equal to  $C-D$.  By definiton, $\tau = \frac{C-D}{m}$, where the   “denominator"  ($m$)  is the total number of all possible pairs combinations, $N(N-1)/2$. Here,  $m  \equiv (C+D) $=$ 26 \times  25/ 2 \equiv  325$.}; the 2-sided $p$-value is equal to 0.00.

\section{ Team Final   Place}    \label{teamfinalplace} 

Similarly to the above, one can define     “best team final place" measures,  such as $P_L^{(\#)}$  together with a  $B_L^{(\#)}$,  based on the final place of  3   “best  riders",   at the end of the race: 
$P_L^{(\#)}$ 
has been defined in Eq.(\ref{PLteameq}); recall that this measure refers to many various riders. In order to adjust  the team ranking by only considering the riders ($j$) who  finish the race, one defines

 \begin{equation}\label{Bteameq}
  B_L^{(\#)}  = \Sigma_{j=1}^{3} \; \; p_{j,L}^{(\#)}
\end{equation}  
as for $  A_L^{(\#)} $, defined in Eq.(\ref{Ateameq}).  Let it be emphasized again that
  in Eq.(\ref{Bteameq}),  $j$ = 1, 2, 3, refers to  3   “best finally placed"  riders, of the team ${(\#)}$ in various stages, but who     {\it have completed all} $L$ stages.

The  $P_L^{(\#)}$   and $B_L^{(\#)}$   results for VSJ are given in Table \ref{Table2PBVSJ}.
Observe at once that the team hierarchy is quite different depending on the measure. The most important scrambling occurs for  the main  (approximatively 10) teams: in particular, COR moves from the $7th$ rank to the $1st$,  TEN gains 3 ranks, but  DSM, the leader according to  $P_L^{(\#)}$, loses one rank, but BOH goes down 3 ranks.

From a statistical difference perspective, the Kendall $\tau$ Rank-Rank correlation coefficient is equal to 0.87692 (Score = 285; Denominator = 325); the 2-sided $p$-value is equal to 0.00.

 Furthermore, notice that  the  $B_L^{(\#)} $ values are always greater than the  $P_L^{(\#)} $ values. This is due to the fact that 
 the finishing riders, after $L$ stages, have not always been involved in the competition for a   “good" place at the  intermediary  stages, while {\it a contrario} riders involved in   “mountain stages" are not those (usually) involved in   “flat stages ending in a sprint".  This again shows that  an excellence overall ranking measure for teams demand a consistent competition by all the riders of a given team.
 
 The role of these riders is emphasized through a discussion of the   “weight" of riders in contributing to a team   “success" in the subsequent section. 
 
 
 \section{Overall Best Riders Place Emphasis} \label{overallbestriderplaceemphasis}

 It has been discussed that the ranking of a team depends on a specific measure based on  a rider place (or time) at the end of a stage. It has been shown how one can rank teams depending on the   “best 3 riders" at the end of a stage.  Usually, 
  at the end of the $L$-stage race,  riders are hierarchically ranked, according to their  aggregated time,  $T_L^{(d)}$, to be distinguished from    $P_L^{(d)}=  \Sigma_{l=1}^L \; \; p_s^{(d)} $, resulting from summing all arriving places of a given $d$ rider.
 
 In fact, one can similarly calculate for each rider   the aggregated final place  after the  $s$-th stage:
  \begin{equation}\label{Psdridereq1}
P_s^{(d)} \;= \;  \Sigma_{l=1}^{s} \;\; p_l^{(d)} \; ,
 \end{equation}
 leading to the $P_L^{(d)}$ at the end of the race.  Thus,  the riders can be ranked in increasing order according to their $P_L^{(d)}$ value, called $g_{j,L}^{(\#)}$. The top (e.g., 26, for saving space) best riders according to their $g_{j,L}^{(\#)}$ are given in Table \ref{Table3dLUXPGVSJ} for illustration.
 
  Interestingly, the   “winner" of the VSJ  race, from such a measure,  should then be Tivani, G.N. (from COR); he arrived  $12th$ in the time ranking.  In contrast, López,  M.Á. (from MED) who had the fastest time for the whole race is ranked $15th$  according to the   $P_L^{(d)}$ measure{{\color{black} \footnote{ {\color{black}This observation allows to refer to Cherchye and Vermeulen (2006) who propose a method combining $place$ $and$ $time$ for emphasizing the truly  best rider for a multi-stage race. Thus, further research can be suggested extending the Cherchye and Vermeulen (2006) approach toward finding the “best team".}}}.
 
 From such a ranking, one can obtain a so called $G_L$  team ranking from the finally placed   “best 3" riders, i.e. 
 
  \begin{equation}\label{Gteameq}
 G_L^{(\#)}\; = \;  \Sigma_{j=1}^{3} \; \; g_{j,L}^{(\#)} \;.
 \end{equation}  
 
 In brief, this   “team  value" results from the standing places of the best placed 3 riders, in each $s$ stage,  but who finished the race.
 
 Moreover, 
 a double summation, on $s$ and on the 3 best  three riders who finished the whole race, leads to 
 
   \begin{equation}\label{SPteameq}
   D_L^{(\#)}\;  =  \Sigma_{i=1}^3  \; \; \Sigma_{l=1}^{s} \;\; p_{j,s}^{(\#)} \; ,
   \end{equation}  
   
 In brief, this   “team  value" results from the aggregation, over all  $L$ stages, of the places of the best placed 3 riders, in a given stage $s$, but for riders who finished the race.

   Finally, for comparison of  place due to time  ranking,  the pertinent  team ranking based on the final place of the three finishing riders  {\it with the best final time}, can be calculated as $U_L^{(\#)}$; these $U_L^{(\#)}$ values and the corresponding  team hierarchy are also given   in columns 4-5 of Table \ref{Table3dLUXPGVSJ}. Notice that one could also calculate the team ranking from the 3 best placed riders in the $P_L^{(d_k)}$ ranking in Table  \ref{Table3dLUXPGVSJ}, i.e.,    $\Sigma_ {k=1}^3 P_{d_k,L}^{(\#)}$, but this is nothing else that $B_L^{(\#)}$ given in Table \ref{Table2PBVSJ}.

        \begin{table}\begin{center} 
\begin{tabular}{|c||c|c||c|c||c|c|c|c|c|c|c|} \hline 
 rank	&$T_L^{(\#)}$	&team      &$[T_L^{(\#)}]^{(*)}$&     team&  $A_L^{(\#)}$ &	team \\ \hline   \hline					
1	&	77:36:22	&	IGD	&	77:05:22	&	IGD &	77:07:53	&	MOV	\\
2	&	77:38:46	&	MOV	&	77:07:42	&	MOV &	77:08:14	&	MED	\\
3	&	77:39:26	&	MED	&	77:08:26	&	MED &	77:08:19	&	IGD	\\
4	&	77:40:27	&	SOQ	&	77:09:27	&	SOQ &	77:12:57	&	TFS	\\
5	&	77:41:43	&	GBF	&	77:10:43	&	GBF 	&	77:14:27	&	AST	\\
6	&	77:42:24	&	AST	&	77:11:24	&	AST 	&	77:17:04	&	SOQ	\\
7	&	77:43:48	&	TEN	&	77:12:48	&	TEN 	&	77:18:03	&	GBF	\\
8	&	77:43:54	&	TFS	&	77:12:54	&	TFS 	&	77:18:59	&	TEN	\\
9	&	77:46:57	&	SEP	&	77:15:57	&	SEP 	&	77:22:07	&	DSM	\\
10	&	77:51:03	&	BOH	&	77:20:03	&	BOH &	77:36:01	&	IPT	\\
11	&	77:52:22	&	DSM	&	77:21:22	&	DSM &	77:38:27	&	COR	\\
12	&	78:01:03	&	COR	&	77:30:03	&	COR &	77:42:08	&	BOH	\\
13	&	78:07:09	&	IPT	&	77:36:09	&	 IPT  &	77:47:28	&	SEP	\\
14	&	78:14:34	&	EOK	&	77:43:34	&	EOK &	77:52:27	&	EOK	\\
15	&	78:27:40	&	AVF	&	77:56:40	&	AVF 	&	78:08:39	&	EGD	\\
16	&	78:28:11	&	ATF	&	77:57:11	&	ATF 	&	78:10:40	&	ATF	\\
17	&	78:33:26	&	EGD	&	78:02:26	&	EGD &	78:15:00	&	AVF	\\
18	&	78:49:49	&	TBG	&	78:15:02	&	CHI 	&	78:23:56	&	CHI	\\
19	&	78:53:19	&	PCV	&	78:18:49	&	TBG 	&	78:30:46	&	PCV	\\
20	&	79:04:23	&	CTQ	&	78:22:19	&	PCV 	&	79:02:34	&	CTQ	\\
21	&	79:30:12	&	CHI	&	78:33:23	&	CTQ &	79:20:32	&	ITA	\\
22	&	79:32:04	&	EMP	&	78:52:42	&	ARG &	79:32:09	&	ARG	\\
23	&	79:47:37	&	ITA	&	79:01:04	&	EMP &	79:32:12	&	MDR	\\
24	&	79:48:37	&	MDR&	79:16:37	&	ITA 	&	79:39:22	&	EMP	\\
25	&	79:57:13	&	ARG	&	79:17:37	&	MDR &	79:40:03	&	URU	\\
26	&	80:05:09	&	URU	&	79:34:09	&	URU &	79:45:54	&	TBG	\\\hline		
	\end{tabular}
\caption{Resulting time ranking of teams  {\it at the end of the 2023 VSJ}, according to the $T_L^{(\#)}$   or  $A_L^{(\#)}$   indicators,  defined in Eq.(\ref{Tteameq}) and Eq.(\ref{Ateameq}), respectively; thus, on one hand, from the (usual) sum of the finishing time of the   “best" 3 riders of the team  {\it after each stage}, and, on the other hand,   “adjusted" in order to be only taking into account those riders who finished the whole race,  respectively. $^{(*)}$ The central $[T_L^{(\#)}]$	data column is the  final team time  officially  reported by the organizers, but including a 31' error: see Appendix  {\color{black} B}.}\label{Table1TAVSJ}
 \end{center}
 \end{table}

   \begin{table}\begin{center} 
\begin{tabular}{|c||c|c||c|c|c||c|c|c|c|c|c|c|} \hline 
 rank	&	$P_L^{(\#)} $ 	&     team& $  B_L^{(\#)} $ &	team 	\\ \hline   \hline			
1	&	409	&	DSM	&	592	&	COR	\\
2	&	430	&	BOH	&	625	&	DSM	\\
3	&	440	&	GBF	&	633	&	TEN	\\
4	&	462	&	MOV	&	635	&	GBF	\\
5	&	471	&	IGD	&	637	&	BOH	\\
6	&	480	&	TEN	&	666	&	MOV	\\
7	&	547	&	COR	&	746	&	IGD	\\
8	&	571	&	SOQ	&	768	&	TFS	\\
9	&	588	&	AST	&	872	&	SOQ	\\
10	&	612	&	TFS	&	879	&	AST	\\
11	&	667	&	IPT	&	899	&	IPT	\\
12	&	882	&	MED	&	956	&	MED	\\
13	&	979	&	SEP	&	1145	&	SEP	\\
14	&	1031	&	EOK	&	1243	&	ATF	\\
15	&	1064	&	TBG	&	1330	&	EGD	\\
16	&	1119	&	ATF	&	1358	&	EOK	\\
17	&	1213	&	EGD	&	1373	&	TBG	\\
18	&	1308	&	AVF	&	1444	&	AVF	\\
19	&	1354	&	PCV	&	1488	&	CHI	\\
20	&	1367	&	CHI	&	1539	&	PCV	\\
21	&	1676	&	ARG	&	1777	&	ARG	\\
22	&	1690	&	CTQ	&	1804	&	CTQ	\\
23	&	1831	&	EMP	&	1892	&	ITA	\\
24	&	1842	&	ITA	&	1947	&	URU	\\
25	&	1880	&	URU	&	1952	&	EMP	\\
26	&	1974	&	MDR	&	2066	&	MDR	\\\hline	
	\end{tabular}
\caption{Resulting place ranking of teams  {\it at the end of the VSJ}, according to the $ P_L^{(\#)} $   or  $ B_L^{(\#)} $   indicators,  Eq.(\ref{PLteameq}) and Eq.(\ref{Bteameq}), respectively; thus, on one hand, from the sum of the finishing place of the   “best" 3 riders of the team  {\it after each stage}, and, on the other hand,   “adjusted" in order to be only taking into account those riders who finished the whole race.} \label{Table2PBVSJ}
 \end{center}
 \end{table}
  
   \begin{table}\begin{center} 
\begin{tabular}{|c||c|c||c|c||c|c|c||c|c|c|c|} 
 \hline 
 rank	&	$D_L^{(\#)}$ &	team& $U_L^{(\#)}$& team&$ P_L^{(d)} $ &  $ rider$  &  team &$G_L^{(\#)} $&team	\\ \hline	\hline		
1	&	28	&	DSM	&	32	&	IGD	&	103	&	TIVANI		&	COR	&	39	&	BOH	\\
2	&	31	&	BOH	&	34	&	MOV	&	139	&	SAGAN		&	TEN	&	40	&	GBF	\\
3	&	32	&	GBF	&	44	&	MED	&	157	&	GAVIRIA		&	MOV	&	40	&	DSM	\\
4	&	42	&	MOV	&	65	&	TFS	&	169	&	SIMMONS	&	TFS	&	42	&	COR	\\
5	&	44	&	IGD	&	67	&	SOQ	&	175	&	NIZZOLO		&	IPT	&	44	&	TEN	\\
6	&	44	&	TEN	&	76	&	AST	&	176	&	ANDRESEN	&	DSM	&	50	&	MOV	\\
7	&	47	&	COR	&	82	&	TEN	&	179	&	VAN POPPEL	&	BOH	&	59	&	IGD	\\
8	&	52	&	SOQ	&	84	&	DSM	&	190	&	ZANONCELLO	&	GBF	&	72	&	TFS	\\
9	&	57	&	AST	&	86	&	GBF	&	194	&	MARKL		&	DSM	&	97	&	SOQ	\\
10	&	63	&	TFS	&	95	&	IPT	&	197	&	MAGLI		&	GBF	&	97	&	IPT	\\
11	&	66	&	IPT	&	109	&	BOH	&	213	&	RIVERA		&	IGD	&	98	&	AST	\\
12	&	87	&	MED	&	111	&	COR	&	213	&	ROJAS		&	CHI	&	110	&	MED	\\
13	&	104	&	SEP	&	116	&	SEP	&	215	&	GANNA		&	IGD	&	140	&	SEP	\\
14	&	106	&	EOK	&	151	&	EOK	&	223	&	HIGUITA		&	BOH	&	166	&	ATF	\\
15	&	106	&	TBG	&	173	&	ATF	&	227	&	LOPEZ		&	MED	&	182	&	EGD	\\
16	&	110	&	ATF	&	180	&	EGD	&	228	&	PAREDES	&	SEP	&	187	&	TBG	\\
17	&	119	&	EGD	&	185	&	AVF	&	232	&	RUBIO		&	MOV	&	191	&	EOK	\\
18	&	121	&	PCV	&	188	&	CHI	&	235	&	BENNETT		&	BOH	&	198	&	AVF	\\
19	&	124	&	AVF	&	221	&	PCV	&	241	&	OSS			&	TEN	&	204	&	CHI	\\
20	&	124	&	CHI	&	236	&	CTQ	&	242	&	VIVIANI		&	COR	&	221	&	PCV	\\
21	&	145	&	ARG	&	284	&	EMP	&	247	&	KONYCHEV	&	COR	&	257	&	CTQ	\\
22	&	156	&	CTQ	&	286	&	URU	&	248	&	BONILLO		&	GBF	&	259	&	ARG	\\
23	&	156	&	ITA	&	295	&	MDR	&	253	&	CRAS		&	TEN	&	277	&	ITA	\\
24	&	160	&	EMP	&	297	&	TBG	&	254	&	TREJADA		&	AST	&	288	&	EMP	\\
25	&	164	&	URU	&	297	&	ITA	&	255	&	WELSFORD 	&	DSM	&	291	&	URU	\\
26	&	169	&	MDR&	315	&	ARG	&	258	&	MESSINEO	&	CTQ	&	306	&	MDR	\\\hline
	\end{tabular}
\caption{Ranking of teams  {\it after  the last VSJ stage},  resulting from the sum of places of a team at the end of each stage $s$, i.e., $D_L^{(\#)}$, Eq. (\ref{SPteameq}). An other ranking in column 9, $G_L^{(\#)}$, i.e., Eq.(\ref{Gteameq}), results from the ranking of the best finishing 3 riders of a team, according to the place they are finishing in each stage $s$. Such top 26 riders, with the sum of their places is given in the central columns, 6-8. The team ranking based on the final place of the three finishing riders  {\it with the best final time},  $U_L^{(\#)}$, is given in columns 4-5.} \label{Table3dLUXPGVSJ}
 \end{center}
 \end{table}

 \section{Discussion}\label{Discussion}
 
  From the present research on, and proposal of, new team ranking  indicators, one can observe:
  on one hand, the numerical data much indicates that the new  indicators bring new quantitative information on the various team's   “values"  at the end of a multi-stage cyclist race. Indeed, it can  be   observed that the new  indicators better distinguish the ranking through the cumulative sums of the places of riders rather than their finishing times.  On the other hand, the  indicators indicate a different team hierarchy if only the finishing riders are considered.
  
  The new ranking proposes less discussion on the {\it ex aequo}s. Nevertheless, it is fair to admit that some {\it ex aequo}s are still possible: indeed,  the measures are based on a finite sum of integers. This is unavoidable, but less often occurring if the number of stages $L$ is large. Moreover, {\it ex aequo}s are less likely if one uses  indicators based on the   “cumulative weight" of riders as in  $P_L^{(\#)}$ and  $B_L^{(\#)}$, as can be  seen in  Table \ref{Table2PBVSJ}.
 
 {\color{black}  A bonus   pertains to the methodology: it is simple;} it starts from downloading the race place (and time)  of each rider after each stage.
  The final result demands to make sums in an appropriate way. There is no 
  {\color{black}   “bonus time"   of the finishing riders before  further ranking, nor the}
    need to invent  complicated arbitrary weights on some inverted finishing race order, as for example done in the   “yellow jersey" in VSJ or the   “green jersey" competition in Tour de France. {\color{black} Moreover to decouple final race measures from  “time or point bonus"   due to intermediary sprints\footnote{The  “time bonus" system was invented for allowing sprinters, who lose much  time in  mountains stages, some incentive expectation in the battle for the final classification.},  allows for another set  of {\color{black} team } strategies, {\color{black}  as those actually selected by riders or coaches}. } 
    {\color{black}  In this respect, i.e.,  considering various team best specificities, thus playing on the possible performance of riders according to the stage type, see Rogge et al. (2012).}
  
Thus, the complete list of riders at the end of each stage can be easily downloaded and stored according to their  arrival time and/or place. Technically speaking, it is sometimes convenient to organize and to store the lists according to the bib of each rider. Even though some algorithm can be invented, some summations are more conveniently done manually.

For a pertinent comparison of   indicators, one has  performed a  Kendall-$\tau$ correlation test, along 
 \url{https://www.wessa.net/rwasp_kendall.wasp\#output}. The latter website   provides the two-sided $p$-value. Moreover, the latter  (free on line) website     provides scatter plots of the $X$ and $Y$ variables and alternatively of their respective ranks. For space saving, these plots are not shown, since they are not carrying any peculiar information of present interest. In all cases, the results are found to be statistically significant.

 Recall that a positive (negative) $\tau$ indicates a so called high (low) rank correlation. It is found that 
  the number of concordant pairs,  i.e.,  when the rank of the second variable $Y$ is greater than the rank of the former variable $X$,  is relatively small., - being even null at the end of this study case. 
  
  {\color{black} Remember that it has been pointed out that a brief set of considerations on improving the Kendall $\tau$ usage, in particular within economic perspectives,  can be found in Appendix A, based on  “weighted preferences'' notions as discussed by Can (2014).}

 \section{Conclusions}\label{Conclusions}
 
 It is commonly admitted that cyclist races are won by one rider, but the role of the team is of crucial importance  (Albert, 1991; {\color{black} Mignot, 2015, 2016}; Cabaud, 2022). In fact, cyclist races are quite different from other sport competitions, emphasising individual athletes.
 Even team competitions, like football (soccer), basketball,  hockey, rowing, etc.,  which sometimes have some focus on specific athletes, or even animals (Lessman et al., 2009) have team quality derived from (integer) numbers, corresponding to some rank and statistics (Anderson, 2015; McHale and Relton, 2018; Kharnat et al., 2020). 
 
  Recall that one aim of the study is to propose an objective   “team value"  measure, with subsequent ordering, whence hierarchy,  for multi-teams competitions. Based on such considerations,  a new, non classical, way of ranking (professional) cyclist teams is proposed.   The numerical analysis is centered on the final hierarchy.  Some illustration is based on a recent race, presenting aspects of more famous multi-stage races, but without loss of generality.  Thereafter, comparing such  new  indicators to the   usual previous one is in order; 
   {\color{black}  one finds distinguishable features, - like clusters, as illustrated in Figs. 1-3, roughly distributed according to the team UCI level, recalled in Table 6.} Thus the linear fits are merely a guide for the eye. 

The main conclusions are, on one hand, 
  that  the new indicators distinguish the  {\color{black} team } ranking through the cumulative sums of the places of riders rather than their finishing times.
   The new ranking proposes less discussion on the {\it ex aequo}s. It is fair to admit that some {\it ex aequo}s are still possible, since the measures are based on a finite sum of integers.
  However, this  occurrence is less likely when considering place-based ranking.  
 
 Moreover, it seems that one can argue that the place accounting should demand more riding action till the end of each stage, thus fuller competition.  A strong sport-based argument in favour of the new  indicators goes as follows indeed: using the place-based  indicators permit to imagine that riders (and team coaches) will have to choose on more different strategies than those existing as of now. Basically, it is expected that riders will attempt to obtain a   “good place", irrespective of their stage time.  Thus, no need to say that if such  indicators had been implemented, in this 2023 VSJ race, the final results might have been different.

  Moreover, these new   indicators can be of interest for betting schemes (Yuce, 2021; Etuk et al., 2022), {\color{black} and/or e-gamers (Beliën et al., 2011).}
  However, this study is not intended to predict the results of a race. 
  
  As other arguments in favour of the  introduction of these new ranking measures, one can consider their interest by sponsors, since the presentation of teams on a podium at the protocol time is the source of a non negligible publicity. 
  Notice that  UCI rules permit 6 distinctive jerseys for leading riders in such multi-stage races.  There does not seem to be a limit for team   “special bibs".  Thus, such  indicators can be implemented, thereby increasing the offer to new sponsorship. For example, by extension of the leading rider jersey notion, one could imagine that one defines a   “green bib for teams", similar to the distinctive  (say, yellow, in Tour de France) bib, for the best time team ranking after $s$ stages. 

 In this line of thought, one should remark that the analysis is based on data for which no strategy was {\it a priori} developed; the role of a coach is not introduced. It should be relevant, and even exciting to see how new strategies will be developed in order to be the most valuable teams along these new operational lines.


{\color{black}  This can be generalized;  “further research" suggestions follow.}
 
 Indeed, notice that one can calculate intermediary indicators values: e.g., one can calculate for all riders ($i$)  their  time (excluding or not time bonuses)  after  $s$ stages
  
 \begin{equation}\label{tsteameqs}
t_{i,s}^{(\#)}  = \Sigma_{l=1}^{s} \;\; t^{(\#)}_{i,l}\;.
\end{equation}

 Thereafter the     “adjusted team  intermediary time", after $s$ stages, - whence after removing running time contributions from  riders not going further than the $s+1$ stage, can be  calculated as follows 
 
 \begin{equation}\label{Asteameq}
  A_s^{(\#)}  = \Sigma_{j=1}^{3} \; \; t_{j,s}^{(\#)}
\end{equation}  
where, in Eq. (\ref{Asteameq}),  $j$ = 1, 2, 3 refers to the 3 riders of the team ${(\#)}$ having the best finishing time after  $s$ stages.

 One can also define,  calculate, for all riders ($i$)  their  place  after  $s$ stages

 \begin{equation}\label{psteameqs}
p_{i,s}^{(\#)}  = \Sigma_{l=1}^{s} \;\; p^{(\#)}_{i,l}\;, 
\end{equation}
and
 \begin{equation}\label{Bsteameq}
  B_s^{(\#)}  = \Sigma_{j=1}^{3} \; \; p_{j,s}^{(\#)}
\end{equation}  
where, in Eq. (\ref{Bsteameq}),  $j$ = 1, 2, 3 refers to the 3 riders of the team ${(\#)}$ having the best finishing places after  $s$ stages. 

Finally, one may hereby   propose further research on  longer (3 weeks) multi-stage races\footnote{{\color{black} At the time of revising this paper, a pertinent example  occurred: a huge set of riders abandoned the (3 week long) Giro d'Italia, - because of Covid constraints. Several of these riders, e.g., Ganna, Evenepoel, Gandin, Vendrame, ..., had been  implied in the first week team standings, whence had much implication on the final (time) ranking. This confirms one of the arguments sustaining the  aim and discussion of this study, i.e., the crucial “value" of the finishing riders in measures.} }. One can likely predict that the  Kendall-$\tau$ values  will tend to become smaller at the end of such races, giving  arguments in favour of the application of the new indicators, whence also leading to  {\color{black} enhanced} competition {\color{black} through new strategies}.


\begin{figure}[ht]   {\color{black} 
\includegraphics[width=0.9\textwidth] {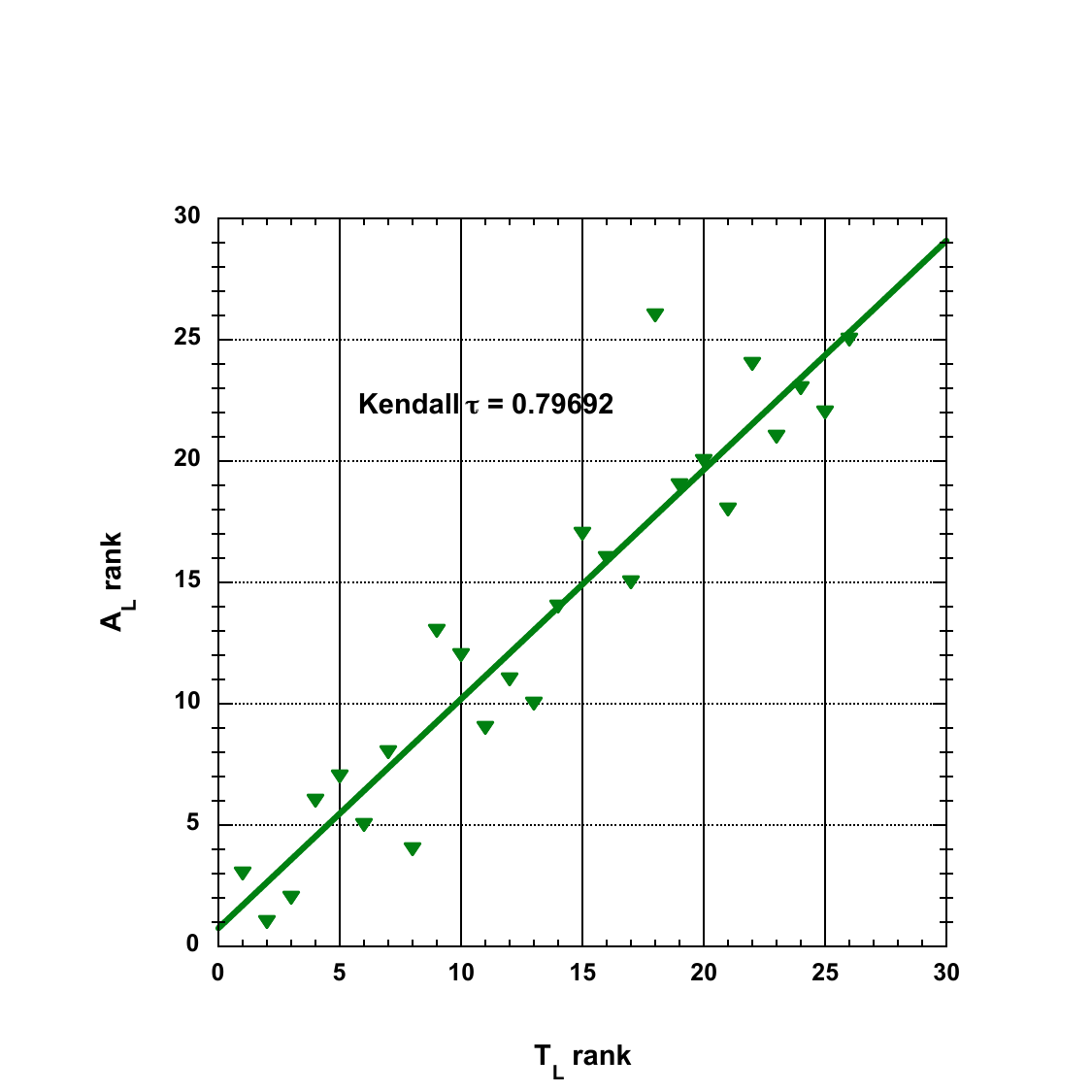} 
\caption{ 
Scatter plot of the rank correlation between  $T_L^{(\#)}$ and $A_L^{(\#)}$,  with mention of the Kendall $\tau$ value ($\simeq 0.7969$); the best linear fit obeys :  y = 0.766154 + 0.943248 \;x, with  $R^2$ $\simeq$ 0.88972. }}
\label{figPlot1TA}
\end{figure}

\begin{figure}[ht]   {\color{black} 
\includegraphics[width=0.9\textwidth] {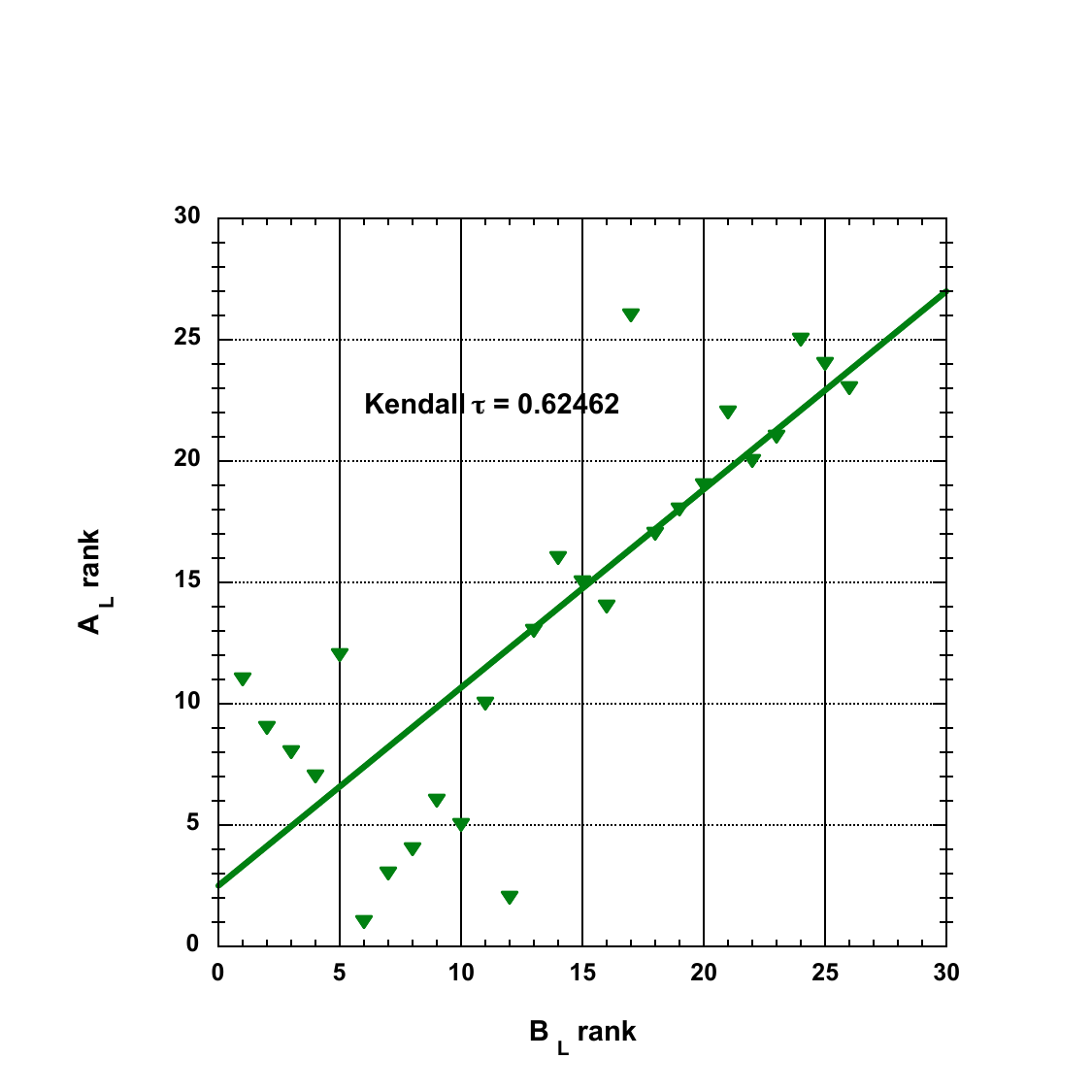} 
\caption{ 
Scatter plot of the rank correlation between  $B_L^{(\#)}$ and $A_L^{(\#)}$,  with mention of the Kendall $\tau$ value ($\simeq 0.6246$); the best linear fit obeys : y = 2.473846 + 0.816752 \;x,   with $R^2$ $\simeq$ 0.66708.
Nevertheless, notice the data points different type of clustering on each side of $r = 13$.}}
\label{figPlot4AB}
\end{figure}
    
\begin{figure}[ht]  {\color{black} 
\includegraphics[width=0.9\textwidth] {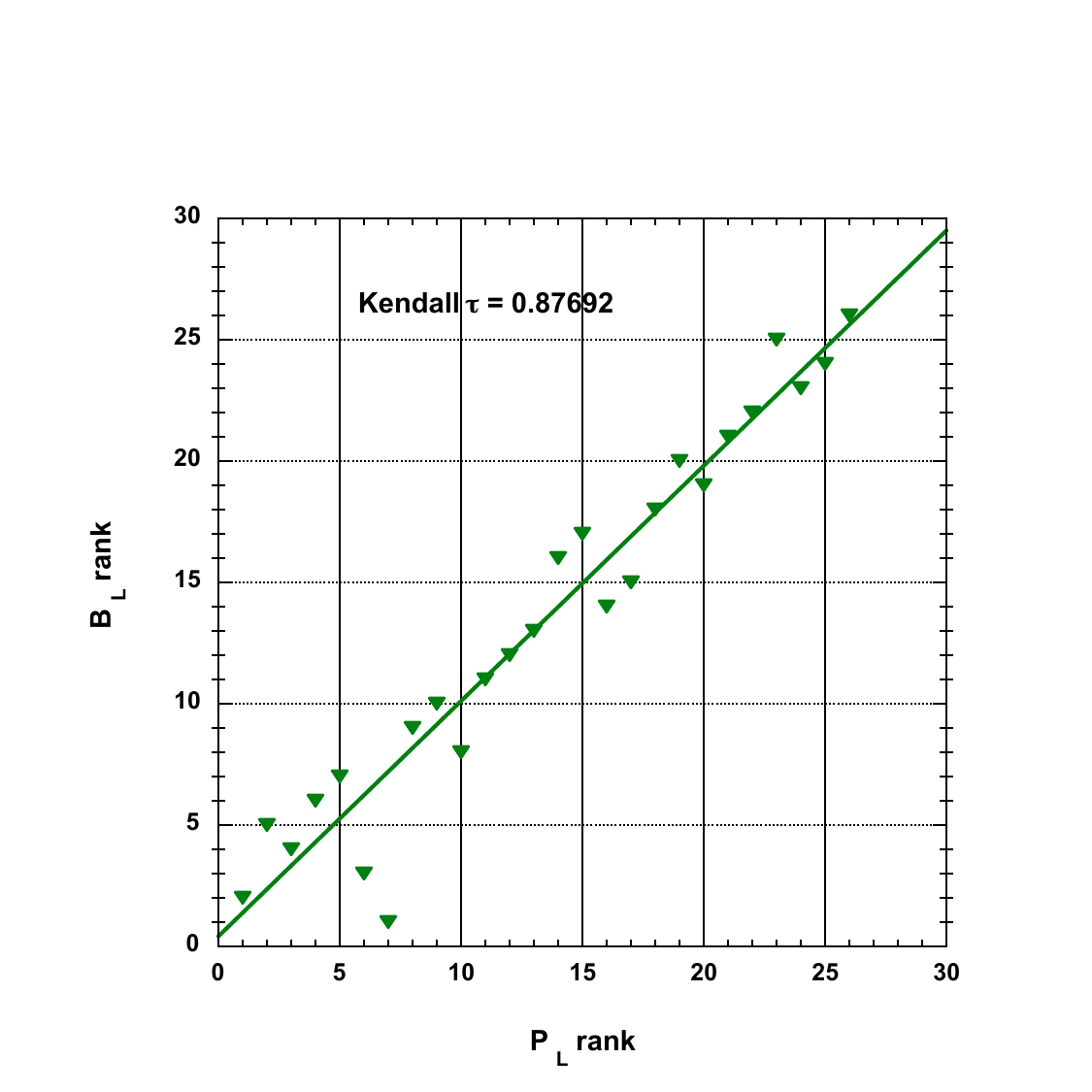} 
\caption{ 
Scatter plot of the  rank correlation between  $B_L^{(\#)}$ and $P_L^{(\#)}$, with mention of the Kendall $\tau$ value ($\simeq 0.8769$); the best linear fit obeys : y = 0.433846 + 0.967863\; x,   with $R^2$ $\simeq$ 0.93676. Notice the data points disordered cluster below $r = 7$.}}
\label{figPlot3BP}
\end{figure}

 \clearpage

   {\bf  {\color{black} Appendix    A: Improving on the Kendall $\tau$ coefficient: the weighted Kemeny distance.} }

   For examples, in sport ranking, in academia ranking, the disagreement at the top or at the bottom of a ranking might have some drastic influence. In sport, the prize money for the first teams (or riders) is quite weighted. The last teams in the ranking face relegation and loss of money contracts. Thus a swap in positions, in both cases, may be very critical. Thus, the score ruling and aggregation schemes are demanded to be robust and statistically significant (Tsakas, 2020).  Therefore, quoting Can (2014) {\it ''it makes sense to assign more dissimilarity (weight) to a change at (...) critical positions.''}
   
 As pointed out by  Can (2014), Csató (2017b),  and/or  a reviewer, the Kendall $\tau$ coefficient does not take into account the precise position of dissimilarities when comparing both linear ranking sets. In particular,  there is neither a discrimination about the rank difference of a pair in both lists, nor  about their relative position in each list. This defect can be (practically) overcome through searching for {\it elementary changes} in both linearly ordered ranking, taking into account all possible item permutations. The permutation number values  can be ranked in a vector form. Next, one considers  sums of  $path$ $distances$ between the elementary changes vector components.  To find the minimum sum  is not a trivial task (Can, 2014); see also  \url{https://ncatlab.org/nlab/show/Cayley+distance}.
 
 

 Considering a debatable potential effect on ranking conclusion emerging from a mere Kendall $\tau$ value, Csató (2017b) proposes to weight the position of discordant pairs  when manipulating the data in order to obtain two identical lists.  He proposes a mere hyperbolic function: $w_C= 1/r$, $r\;\in\;[1,r_{M}-1]$, based on the lowest rank $r$   of an item of a discordant  pair. Conforming to Csató, one can examine if such a weight has an effect when measuring the discrepancy between the resulting  indicators discussed the main text. The classical Kendall coefficient $\tau$ would correspond to choose a $w_K=1$ for all permutations. For some scientific addition, the permutation values are also weighted through another weight distribution:  $w_A= \sqrt{1/r}$, smoother  than  $w_C$, as shown in Fig. \ref{Plot9weightswKwAwCdiscrete}, for the particular case $r_M =26$. Thus, one can classically measure the number of discrepancy pairs  $D$, obtain the “score" $S$, thereafter the Kendall $\tau$ coefficient from $S/m$, - as recalled in a footnote here above, whence the Kemeny distance\footnote{The Kemeny distance is usually equivalent to the Kendall distance (Kemeny and Snell, 1962; Heiser and D'Ambrosio, 2013)} $K= N(N-1)(1-\tau)/4$.                  The procedure can be repeated, appropriately weighting the various swaps, whence  obtaining a “weighted Kemeny distance'' 
  between pairs, 
  called $A$ and $C$. 
 
 
   \begin{table}\begin{center}  
\begin{tabular}{|c||c|c|c|c||c|c|c|c|c|c|} 
 \hline 
 &	$D$ &	$S$ 	&	$\tau$	& $K$ & $P$ 	&   $A$    &   $ C$ 	&$Q$	\\ \hline	\hline		
$A_L,\;[T_L]$	&	29	&	267	&	0.82154	&	29	&	101	&	29.136	&	8.9667&27		\\
$A_L,\;P_L$	&	59	&	207	&	0.63692	&	59	&	103	&	31.531	&	10.493&59	\\
$A_L,\;T_L$	&	33	&	259	&	0.79692	&	33	&	131	&	38.655	&	12.371&33		\\
$A_L,\;B_L$	&	61	&	203	&	0.62462	&	61	&	155	&	46.351	&	15.724&61		\\
$P_L,\;[T_L]$	&	46	&	233	&	0.71692	&	46	&	152	&	43.029	&	13.199&45	\\
$T_L,\;[T_L]$	&	6	&	313	&	0.96308	&	6	&	56	&	17.869	&	6.2649&7		\\
$B_L,\;[T_L]$	&	56	&	213	&	0.65538	&	56	&	126	&	38.912	&	13.773&57		\\
$P_L,\;T_L$	&	48	&	229	&	0.70462	&	48	&	178	&	49.409	&	14.790&48		\\
$P_L,\;B_L$	&	20	&	285	&	0.87692	&	20	&	146	&	42.726	&	14.199&24		\\
$B_L,\;T_L$	&	60	&	205	&	0.63077	&	60	&	140	&	42.860	&	14.887&60		\\
 \hline
	\end{tabular}
\caption{Characteristics values leading to measures of the “Kemeny distance'' (Kendall $\tau$  without normalisation) between pairs of  indicators, for a few cases examined in the main text: 
 number of discordant pairs $D$; score $S$; a number of permutations $P$ or $Q$ depending on the chosen vector space (see text); Kendall $\tau$ coefficient; Kemeny  distance $K (\equiv D)$ (\url{https://ncatlab.org/nlab/show/Kendall+tau+distance}). The weighted Kemeny distance with either square root or mere (-1) hyperbolic weight distribution on permutations is $A$ or $C$, respectively.
} \label{TableKCAAppA}
 \end{center}
 \end{table}

 This has been done for a few relevant  indicators, 
 i.e., measuring the correlation between pairs of  indicators due to the number of discordant and concordant pairs (of teams) in Tables 1-3.
 Not all  indicators of the main text are considered, because a few ($D_L$, $G_L$, and $U_L$) contain {\it ex aequo}s,  leading to practical complications, the solution of which would carry too far from the main points. The results are gathered in Table \ref{TableKCAAppA}.  
 
 Notice that practically, in order to compare the ranks of pairs of teams, it is first useful to organize the teams in alphabetical order, giving them the appropriate rank for a given indicator, as summarized in Table \ref{Tablealphabetrank}. Let it be also observed that the number of permutations in order to reconcile two vectors, or, in other words, measuring their  distance depends on the order of the axes in the chosen coordinate space. Thus, the number of permutations $P$ starting from the ranking, e.g., in Table  \ref{Tablealphabetrank} differs from the corresponding number of permutations $Q$ to be performed in Tables 1-3; necessarily $P \ge Q$.   Alas, the path minimizing metric in order to reconcile two lists is not trivial to find (Can, 2014). The  “winners' decomposition", in Can (2014) wording is chosen to count the numbers of permutations. 
 Notice that  the   indicators distances are also necessarily ordered: $K \ge A \ge C$.

These considerations should be further pursued. Indeed, such ranks can be considered as network node degrees, thereby leading to identifying key nodes in a network (Csató, 2017a), - i.e., their $centrality$ (Rotundo, 2011; Negahban et al., 2016; Yazidi et al., 2022). Moreover, the search for the path minimizing function (Can, 2014) is related to searching for specific paths and communities on networks (Festa and Resende, 2009; Varela et al., 2015; Winston and Goldberg, 2022). This is of much interest when communities, whence potential hierarchy, emerge
 (Eckert and McConnell-Ginet, 1992; Cohendet et al., 2004; Demil and Lecocq, 2006; Yearworth and White, 2018; Wu et al., 2019).

   \begin{table}\begin{center}   \fontsize{8pt}{8pt}\selectfont
\begin{tabular}{|c||c|c|c|c|c|c|c|c|c|c|} 
 \hline 
&\multicolumn{5}{|c|}{rank in indicators}    \\	\hline
$team$ &	$A_L$ &	$[T_L]$ 	&	$P_L$	& $T_L$ & $B_L$ 	\\ \hline	\hline	

ATF	&	16	&	16	&	16	&	16	&	14	\\
ARG	&	22	&	22	&	21	&	25	&	21	\\
AST	&	5	&	6	&	9	&	6	&	10	\\
AVF	&	17	&	15	&	18	&	15	&	18	\\
BOH	&	12	&	10	&	2	&	10	&	5	\\
CHI	&	18	&	18	&	20	&	21	&	19	\\
COR	&	11	&	12	&	7	&	12	&	1	\\
CTQ	&	20	&	21	&	22	&	20	&	22	\\
DSM	&	9	&	11	&	1	&	11	&	2	\\
EGD	&	15	&	17	&	17	&	17	&	15	\\
EMP	&	24	&	23	&	23	&	22	&	25	\\
EOK	&	14	&	14	&	14	&	14	&	16	\\
GBF	&	7	&	5	&	3	&	5	&	4	\\
IGD	&	3	&	1	&	5	&	1	&	7	\\
IPT	&	10	&	13	&	11	&	13	&	11	\\
ITA	&	21	&	24	&	24	&	23	&	23	\\
MDR	&	23	&	25	&	26	&	24	&	26	\\
MED	&	2	&	3	&	12	&	3	&	12	\\
MOV	&	1	&	2	&	4	&	2	&	6	\\
PCV	&	19	&	20	&	19	&	19	&	20	\\
SEP	&	13	&	9	&	13	&	9	&	13	\\
SOQ	&	6	&	4	&	8	&	4	&	9	\\
TBG	&	26	&	19	&	15	&	18	&	17	\\
TEN	&	8	&	7	&	6	&	7	&	3	\\
TFS	&	4	&	8	&	10	&	8	&	8	\\
URU	&	25	&	26	&	25	&	26	&	24	\\
   \hline
	\end{tabular}
\caption{Alphabetical order of teams with their rank in the studied indicators.
} \label{Tablealphabetrank}
 \end{center}
 \end{table}
 
   \begin{figure}[ht]   
\includegraphics[width=0.9\textwidth]{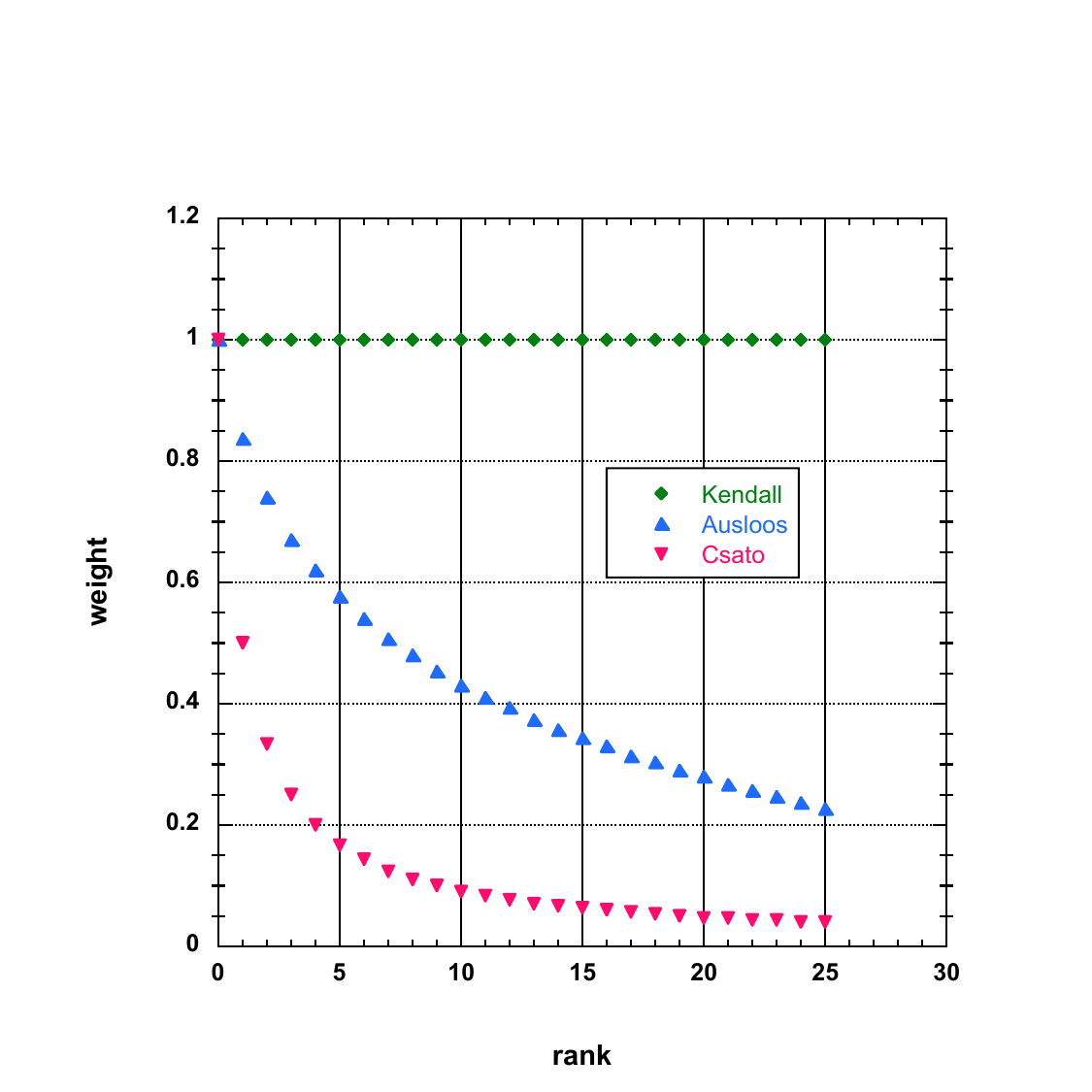}   
\caption{Plot of the weight values distribution used for comparing pair dissimilarities in the  indicators discussed in the main text:  Csató, Ausloos, and Kendall respectively: $w_C\; =1/r$, $w_A\; = \sqrt{1/r}$, $w_K=1$, where $r$ is the lowest rank of a member of the discordant pair.}
\label{Plot9weightswKwAwCdiscrete}
\end{figure}

\clearpage
   {\bf Appendix   {\color{black} B}: Anomalies in  2023 VSJ data reporting} }
   
      \bigskip
   The official data source   
    \url{https://www.vueltaasanjuan.org/clasificaciones/} contains
   several   “errors"  (or    “misprints", or   “confusions"). In particular, 
 
  \begin{itemize}
  \item  
 The finishing time of the 1st stage winner was (is)   reported to be 3h 19' 36".
 
 
Since the  first 85 riders from the main bunch finished the 1st stage in 3h 19' 36", together,  - 16 teams with more than 3 riders in the bunch, those teams supposedly final time   should  be  (3 $\times$  3h 19' 36" =) 9h 58' 48", obviously. 
 
 However, it was  reported that the first 16 teams are supposed to have finished in 9h 57' 48", on stage 1, leading to a 1' error.
   

   \item
      For the final stage (stage 7),   the winner supposedly  rode the stage in  2h 33' 41''. 
 
 Since  three relevant riders for 22 teams finished in a bunch together with the winning rider, their  team finishing time should be  (2h 33' 41'' $\times$ 3 =) 7h 41' 03''. 
 
 However, the organisers report these  team finishing  times to be  7h 11' 03'', thereby missing 30'. 
    
  \end{itemize}
    
    
    Thus, the final team stage   “overall time"  becomes shorter by 31', for the main teams. For completeness, the official time is given in Table \ref{Table1TAVSJ},  as $[T_L^{(\#)}]^*$.
     Fortunately,   these errors do not carry over on the place ranking relative values.
     {\color{black}  For completeness, the used data in the main text, for each stage, is found in Supplemental Materials.}

  \bigskip 
{\bf  Acknowledgements :}
 
   {\it to conserve anonymity, in the peer review process, no acknowledgement is hereby presented; it will await publication time ; reviewers, editor, and private communication expert will be mentioned }
  
{\bf   Data availability} : {\it data is freely available,  see text.} 

    
  {\bf   Funding} : {\it none.}
    
      
      {\bf  Disclosure Statement  on competing interest} : {\it Neither  relevant financial nor non-financial  competing interest has to be mentioned.}

 \begin{table}\begin{center}  
\begin{tabular}{|c||c|c||c|c||c||c|c||c|c|c|c|} \hline 	
UCI	&	 Team Sponsor&	country		\\ \hline \hline
acronym	&	World Teams (WTT)	&		acronym	\\ \hline
AST	&	ASTANA QAZAQSTAN TEAM	&	KAZ	\\
BOH	&	BORA - HANSGROHE	&	DEU	\\
DSM	&	TEAM DSM	&	NLD		\\
IGD	&	INEOS GRENADIERS	&	GBR		\\
MOV	&	MOVISTAR TEAM	&	ESP		\\
SOQ	&	SOUDAL QUICK-STEP	&	BEL		\\
TFS	&	TREK-SEGAFREDO	&	USA	\\ \hline \hline
	&	Pro Teams (PRT)	&				\\ \hline
COR	&	TEAM CORRATEC	&	ITA		\\
EOK	&	EOLO-KOMETA CYCLING TEAM	&	ITA		\\
GBF	&	GREEN PROJECT-BARDIANICSF-FAIZANE'	&	ITA	\\
IPT	&	ISRAEL - PREMIER TECH	&	ISR		\\
TEN	&	TOTAL ENERGIES	&	FRA	\\ \hline \hline
	&	Continental Teams (CTM)	&				\\ \hline
ATF	&	AP HOTELS \& RESORTS / TAVIRA / SC FARENSE	&	PRT	\\
AVF	&	AV FATIMA-SAN JUAN BIKER MOTOS-ELECTRO 3	&	ARG		\\
CTQ	&	CHIMBAS TE QUIERO	&	ARG		\\
EGD	&	GREMIOS POR EL DEPORTE-YACO	&	ARG	\\
EMP	&	EC MUNICIPALIDAD DE POCITO	&	ARG		\\
MDR	&	MUNICIPALIDAD DE RAWSON	&	ARG	\\
MED	&	TEAM MEDELLIN-EPM	&	COL		\\
PCV	&	PANAMA ES CULTURA Y VALORES	&	PAN 	\\
SEP	&	SINDICATO DE EMPLEADOS PÚBLICOS DE SAN JUAN	&	ARG	 	\\
TBG	&	TEAM BANCO GUAYAQUIL	&	ECU	 		\\ \hline \hline
&	National Teams (NTM)				&	\\ \hline
ARG	&	ARGENTINA	&	ARG			\\
CHI	&	CHILE		&CHL		\\
ITA	&	ITALY		&ITA			\\
URU	&	URUGUAY	&	URY			\\ \hline
	\end{tabular}
	{\color{black} 
\caption{ UCI  acronym, team sponsors,  and (3-letter country abbreviation ISO-3166-1 ALPHA-3 conventional notation)  country registration for competing teams in the 2023 VSJ race; the conventional UCI levels are distinguished; the alphabetical order of UCI acronyms is used. } \label{TableVSJacron}}
 \end{center}
 \end{table}



 \end{document}